\documentclass[pra,twocolumn,showpacs]{revtex4}
\usepackage{graphicx}
\begin{document}

\bibliographystyle{unsrt}

%%%%%%%%%%%%%%%%%%%%%%%%%%%%%%%%%%%%%%%%%
\title{Compensating for Beamsplitter Asymmetries in Quantum Interference Experiments}
\author{J.L. Liang and T.B. Pittman}
\affiliation{Physics Department, University of Maryland Baltimore County, Baltimore, MD 21250}
%%%%%%%%%%%%%%%%%%%%%%%%%%%%%%%%%%%%%%%%%

%%%%%%%%%%%%%%%%%%%%%%%%%%%%%%%%%%%%%%%%%
\begin{abstract}
The visibility of the quantum interference ``dip'' seen in the Hong-Ou-Mandel experiment is optimized when a symmetric 50/50 beamsplitter is used in the interferometer.  Here we show that the reduction in visibility caused by an asymmetric beamsplitter can be compensated by manipulating the polarization states of the two input photons. We experimentally demonstrate this by using a highly asymmetric 10/90 beamsplitter, and converting an initial dip visibility of $22\%$ to a compensated value of $99\%$.
\end{abstract}
%%%%%%%%%%%%%%%%%%%%%%%%%%%%%%%%%%%%%%%%%


\maketitle

The nonclassical interference effects that can arise when two single-photons are incident on a beamsplitter form the basis of a number of benchmark experiments in quantum optics \cite{fearn89}. When the beamsplitter is perfectly symmetric (eg. 50/50), destructive interference between the two-photon amplitudes associated with both photons being reflected ($A_{rr}$) and both photons being transmitted ($A_{tt}$) leads to the well-known Hong-Ou-Mandel (HOM)``dip'' in the coincidence counting rate between single-photon detectors in the output ports \cite{hong87,shih88}. However, if the beamsplitter is not perfectly symmetric, the visibility of the HOM dip is reduced because $A_{rr}$ and $A_{tt}$ are no longer equally weighted \cite{matthews09}. As the beamsplitter asymmetry becomes larger and larger, more two-photon ``which path'' information is available, and the quantum interference effect is correspondingly diminished.

Somewhat surprisingly, this reduced interference has been found to be extremely useful in a number of quantum information processing applications (QIP) including linear optics quantum computing gates \cite{ralph01,hofmann02,obrien03,sanaka04}, quantum cloning machines \cite{filip04, zhao05,bartuskova07}, and Fock-state filters \cite{sanaka05,sanaka06}. Asymmetric beamsplitters have also been shown to be useful in multi-photon quantum interference experiments \cite{fiurasek02,wang05,liu07,resch07,liu08,ou08a,ou08b}. Because of the growing importance of asymmetric beamsplitters, it is important to fully explore the interference effects associated with them. In this brief paper, we show that the effects of beamsplitter asymmetries in basic two-photon interference experiments can be compensated by manipulating the polarization states of the incident photons. This allows a recovery of indistinguishability and, consequently, a recovery of high visibility dips in the coincidence counting rates. We experimentally demonstrate this idea by using photon pairs from a parametric down-conversion (PDC) source, and a highly asymmetric fiber-based beamsplitter with a 10/90 intensity-splitting ratio.

\begin{figure}[b]
\includegraphics[width=2.25in]{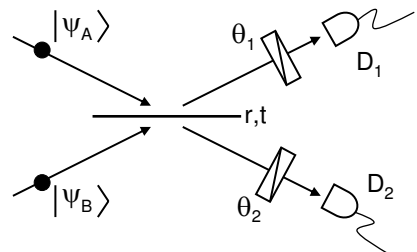}
\vspace*{-.25in}
\caption{Overview of a method to compensate for beamsplitter asymmetries in basic two-photon interference ``dip'' experiments. Unequal reflection and transmission coefficients $r$ and $t$ provide two-photon ``which-path'' information which reduces the visibility of the interference effect. By manipulating the polarization states of the two incident photons, 100\% visibility can be restored.}
\label{fig:overview}
\end{figure}

An overview of the compensation technique is shown in Figure \ref{fig:overview}. Two single-photons are incident on a non-polarizing beamsplitter with amplitude reflection/transmission coefficients $r$ and $t$, and single-photon detectors $D_{1}$ and $D_{2}$ in the output ports. If the polarization states of the input photons (denoted by $|\psi_{A}\rangle$ and $|\psi_{B}\rangle$) are the same, the visibility of the HOM dip is given by  $V=2RT/(R^{2}+T^2)$, where $R\equiv |r|^{2}$ and $T\equiv |t|^2$ are the intensity-splitting coefficients ($R+T=1$). The visibility is 100\% for a 50/50 beamsplitter but, for example, falls to only 22\% for an asymmetric 10/90 beamsplitter because the two incident photons are much more likely to be transmitted than reflected.

The basic idea is to compensate for the beamsplitter asymmetry by using the polarization degree of freedom to reduce the overall magnitude of $A_{tt}$ until it equals that of $A_{rr}$. As shown in Figure \ref{fig:overview}, this is accomplished by inserting polarizers $\theta_{1}$ and $\theta_{2}$ in the output ports, and preparing the appropriate input polarization states $|\psi_{A}\rangle$ and $|\psi_{B}\rangle$. For simplicity, we'll consider linear polarization states, and always align $\theta_{1}$ in the direction of $|\psi_{A}\rangle$, and $\theta_{2}$ in the direction of $|\psi_{B}\rangle$. This keeps the weaker two-photon amplitude $A_{rr}$ at its maximum possible value, but reduces the magnitude of $A_{tt}$ as needed. In this case, the problem can be simplified to depend on a single polarization variable defined as the difference of the initial polarization states, $\Delta\theta \equiv \theta_{1} - \theta_{2}$, and the visibility of the coincidence dip can be shown to be:

 \begin{equation}
 V=\frac{2RT Cos^{2}(\Delta\theta)}{R^{2}+T^2 Cos^{4}(\Delta\theta)}
 \label{eq:dipvis}
 \end{equation}

For the case of the 10/90 asymmetric beamsplitter, the original 22\% dip visibility is increased to 100\% when  $\Delta\theta$ is set to $70.5^{o}$.

%%%%%%%%%%%%%%%%%%%%%%%%%%%%%%%%%%
%%%%%%%%%%%%%%%%%%%%%%%%%%%%%%%%%%%

\begin{figure}[b]
\includegraphics[width=3.5in]{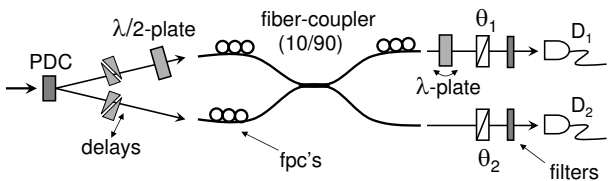}
\caption{Schematic of the apparatus used to demonstrate compensation of beamsplitter asymmetries in two-photon interference experiments. A PDC source is used to send two single-photons into an asymmetric 10/90 single-mode fiber-coupler. Nonclassical ``dips'' in the coincidence count rate between detectors $D_{1}$ and $D_{2}$ are measured as a function of the relative delay of the input photons. A half waveplate ($\lambda/2$-plate) and polarizers are used to perform the compensation technique. Fiber polarization controllers (fpc's) and an auxiliary waveplate ($\lambda$-plate) are used to negate birefringence in the fiber-coupler.}
\label{fig:experiment}
\end{figure}

Figure \ref{fig:experiment} shows the experimental apparatus used to test the compensation technique. A continuous-wave ultraviolet diode laser ($\sim$10 mW at 407 nm) was used to pump a 0.7 mm thick BBO crystal cut for type-I degenerate parametric down-conversion (PDC). The PDC source produced pairs of horizontally polarized photons at 814 nm, which were coupled into a single-mode fiber-coupler with a 10/90 beamsplitter ratio. The output photons were sent back into free-space and detected by single-photon detectors $D_{1}$ and $D_{2}$, which were preceded by interference filters centered near 814 nm with a FWHM bandpass of 10 nm. The nonclassical ``dips'' in the coincidence counting rate (between $D_{1}$ and $D_{2}$) were recorded as a function of the relative optical delay between the PDC photons, which was controlled by translatable glass wedges placed in the input beams.

In order to specify $\Delta\theta$ of equation (\ref{eq:dipvis}), the input state $|\psi_{B}\rangle$ was kept horizontally polarized, and the state $|\psi_{A}\rangle$ was prepared using a half-wave plate to rotate the polarization of the PDC photon in the upper beam. Correspondingly, the polarizer $\theta_{2}$ was always kept at $0^{o}$ (eg. horizontal) and $\theta_{1}$ was always set to match the state $|\psi_{A}\rangle$.

Keeping $|\psi_{B}\rangle$ fixed at $0^{o}$ was not necessary, but simplified the number of elements needed to overcome the deleterious effects of stress-induced birefringence in the fiber-coupler. The alignment procedure involved sending an auxiliary laser beam at 803 nm (not shown) along the PDC paths and into the fiber coupler, and measuring the output polarizations with $\theta_{1}$ and $\theta_{2}$. A combination of three ``paddle-wheel'' style fiber polarization controllers (fpc's) and an extra waveplate in the upper output port could be sequentially adjusted to perform adequate birefringence cancelation \cite{803vs814}.

%%%%%%%%%%%%%%%%%%%%%%%%%%%%%%%%%%
%%%%%%%%%%%%%%%%%%%%%%%%%%%%%%%%%%%

The fpc's were first used to ensure that horizontal and vertical input polarization states emerged intact. The waveplate in the output port (with its fast-axis horizontal) could then be slightly twisted to cancel residual phase-shifts between the horizontal and vertical polarization components, thereby ensuring that arbitrary linear input states $|\psi_{A}\rangle$ would faithfully emerge from the upper port. By keeping $|\psi_{B}\rangle$ horizontal (and consequently $\theta_{2}$ at $0^{o}$) in the experiments,  it was not necessary to cancel these kinds of phase shifts in the lower output path because only the horizontal component of $|\psi_{A}\rangle$ would ever be detected by $D_{2}$. The use of arbitrary input states $|\psi_{B}\rangle$ would have necessitated additional birefringence control in the system.

\begin{figure}[t]
\includegraphics[width=3.25in]{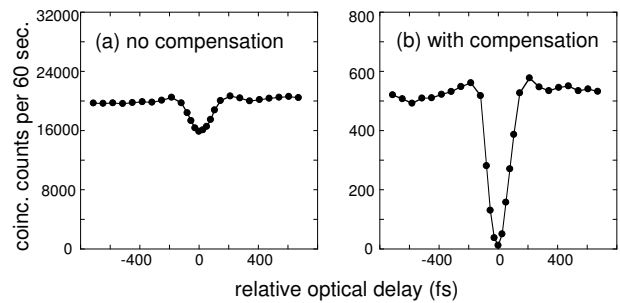}
\caption{Experimental results demonstrating compensation for an asymmetric 10/90 beamsplitter. Figure (a) shows the expected HOM-like ``dip'' with a low visibility of $(21.7 \pm 0.4)\%$. In Figure (b) the visibility is increased to $(98.6 \pm 0.7)\%$ by using the compensation technique. In each plot, the lines simply connect the data points, and the visibility is obtained by least-squares fitting of the data with a simple Gaussian function.}
\label{fig:dipdata}
\end{figure}

Figure \ref{fig:dipdata} shows experimental results demonstrating compensation for the asymmetric 10/90 beamplitter. For Figure \ref{fig:dipdata}(a) the input states and polarizers were all set to horizontal, which corresponds to the case of no compensation. As expected, we see a simple HOM-like ``dip'' in the coincidence counting rate with a limited visibility of only $(21.7 \pm 0.4)\%$. This agrees with the predicted value of $22\%$ \cite{matthews09}.  For the data shown in Figure \ref{fig:dipdata}(b), we adjusted $|\psi_{1}\rangle$ and $\theta_{1}$ so that $\Delta\theta$ was set to the optimal value of $\sim 70.5^{o}$. This performs full compensation, and we see an increase of the dip visibility to $(98.6 \pm 0.7)\%$, which is very close to the quantum mechanical prediction of $100\%$.

%%%%%%%%%%%%%%%%%%%%%%%%%%%%%%%%%%
%%%%%%%%%%%%%%%%%%%%%%%%%%%%%%%%%%%

\begin{figure}[t]
\includegraphics[width=2.75in]{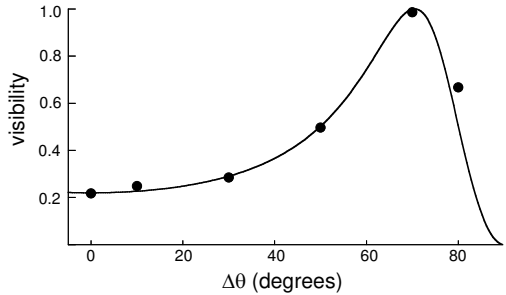}
\caption{Demonstration of the dependence of the dip visibility on the difference between the input polarization states $\Delta\theta$. The solid line is a theoretical plot of equation (\protect\ref{eq:dipvis}) for the beamsplitter with $R=0.1$ and $T=0.9$. The data points were obtained by fitting plots similar to those shown in Figure \protect\ref{fig:dipdata} (the error bars are comparable to the point size). }
\label{fig:visibilitydata}
\end{figure}

In order to further verify equation (\ref{eq:dipvis}), we measured the visibility of ``dips'' analogous to those shown in Figure \ref{fig:dipdata} for several other values of $\Delta\theta$. The results are shown in Figure \ref{fig:visibilitydata}. The close agreement of the data with the theoretical prediction of equation (\ref{eq:dipvis}) using a highly asymmetric 10/90 beamsplitter clearly shows the ability for compensation in two-photon interference experiments.

%%%%%%%%%%%%%%%%%%%%%%%%%%%%%%%%%%
%%%%%%%%%%%%%%%%%%%%%%%%%%%%%%%%%%%

From Figure \ref{fig:visibilitydata} we see that as $\Delta\theta$ is increased from $0^{o}$ towards the optimal value of $70.5^{o}$ the visibility of the dip rapidly increases, but it should be noted that the price paid is a reduction in the overall coincidence rate which rapidly drops as $R_{c} = R^{2}+T^{2}Cos^{4}(\Delta\theta)$.  This is simply due to the fact that as $\Delta\theta$ is increased the overall magnitude of $A_{tt}$ is further reduced until it finally matches that of the much weaker $A_{rr}$. If $\Delta\theta$ is increased past the optimal value, both the counting rate and the visibility drop. These effects become less dramatic as the asymmetry of the beamsplitter is reduced towards the symmetric 50/50 case.

It is important to note that the interference effects in the fully compensated case involve polarization, and are therefore more closely related to the Shih-Alley interferometer \cite{shih88} than the HOM interferometer \cite{hong87}. In other words, the recovered high visibility dip is not simply due to ``both photons leaving the same port'', but involves a suppression of coincidence counts that relies on absorption at the polarizers.  The near $100\%$ dip observed in Figure \ref{fig:dipdata}(b) is due to the fact that the overall two-photon amplitudes $A_{rr}$ and $A_{tt}$ (which include passage through the polarizers) have now become equally weighted. Indeed, the destructive interference ``dips'' can be converted to constructive ``peaks'' by manipulating the overall sign of $A_{tt}$ compared to $A_{rr}$ after the beamsplitter. We verified that this can be done, for example, by choosing appropriate orientations of the polarizers.

In summary, we have experimentally shown that the polarization degree of freedom can be exploited to compensate for beamsplitter asymmetries in simple two-photon interference experiments. This raises the question of experimental arrangements that would be needed to demonstrate analogous compensation techniques based on other degrees of freedom, or related compensation effects for more specialized beamsplitters with polarization-dependent asymmetries \cite{langford05,kiesel05,okamoto05,cernoch06,soubusta07,soubusta08}.

We acknowledge useful discussions with J.D. Franson.  This work was supported in part by part by  the National Science Foundation under grant No. 0652560, and the Intelligence Advanced Research Projects Activity (IARPA) under Army Research Office (ARO) contract W911NF-05-1-0397.

%%%%%%%%%%%%%%%%%%%%%%%%%%%%%%%

%%%%%%%%%%%%%%%%%%%%%%%%%%%%%%%%


\begin{thebibliography}{50}
\bibitem{803vs814} Once optimized, we found the overall birefringence of the system to be fairly insensitive to the wavelength difference between the 803 nm beam and the 814 nm PDC signal, and typically obtained better results after performing the calibration with the much stronger 803 nm beam.
\end{thebibliography}
\end{document}